\documentclass[pre,twocolumn,showpacs]{revtex4}

\usepackage{graphicx}
\usepackage{bm}

\begin{document}

\title{A new mechanism for granular segregation}

\author{D. C. Rapaport}
\email{rapaport@mail.biu.ac.il}
\affiliation{Physics Department, Bar-Ilan University, Ramat-Gan 52900, Israel}

\date{\today}

\begin{abstract}

A novel process is described that produces horizontal size segregation in a
vertically vibrated layer of granular material. The behavior is a consequence
of two distinct phenomena that are unique to excited granular media: vibration
which causes the large particles to rise to the top of the layer, and a
vibrating base with a sawtooth surface profile which can produce stratified
flows in opposite directions at different heights within the layer. The result
of combining these effects is that large and small particles are horizontally
driven in opposite directions. The observations reported here are based on
computer simulations of granular models in two and three dimensions.

\end{abstract}

\pacs{45.70.Mg, 02.70.Ns}

\maketitle

Granular media have been said to embody properties of solids, liquids and
gases, but in actual fact studies of the granular state continue to produce
surprising results that are unique to this class of matter
\cite{bar94,her95,jae96}. Computer modeling plays an important role in these
developments, due to the fact that once it has demonstrated an ability to
reproduce the observed behavior it can be used in trying to understand the
mechanisms involved; examples of the dynamics of granular media that have been
examined by this kind of approach include vertical size segregation by both
vibration \cite{ros87,gal96} and shear \cite{wal83,hir97}, and
vibration-induced surface waves \cite{cle96,lud96,rap98,biz98}.

A further instance of unusual behavior in granular matter that has come to
light very recently involves a granular layer that is vibrated vertically by a
base whose surface profile consists of sawtooth-like grooves (in the
experimental realization the material is contained in a narrow annular region
between two upright cylinders and is thus practically two-dimensional). What is
observed \cite{far99}, both experimentally and in computer simulation, is the
occurrence of horizontal flow. Moreover, the flow direction and magnitude
depend on many of the parameters required to specify the system in an
apparently complex manner. Subsequent more detailed simulation of this
phenomenon \cite{lev01} revealed that the induced flow rate actually varies
with height within the granular layer; even more surprising is the observation
that oppositely directed flows can exist simultaneously at different levels in
the layer. Once the strongly stratified nature of the horizontal flow is
appreciated, the sensitive parameter dependence of the overall horizontal
motion of the layer is readily understood as being a consequence of the
competition between opposing stratified flows.

A combination of two of the phenomena that are unique to granular matter,
namely the already familiar vertical size segregation under vibration, the
so-called ``Brazil nut'' effect, and the recently observed stratified flow
outlined above, suggests an entirely new mechanism for separating the
components of a granular mixture according to particle size. If, under suitable
vibration conditions, the upper and lower layers of the material are able to
move horizontally in opposite directions under the influence of a
sawtooth-shaped base, and the same vibrations cause the large particles to
climb towards the top of the layer, then it is apparent that some degree of
horizontal segregation of large and small particles should occur as a
consequence. A process based on behavior of this kind could well be significant
from an industrial perspective. In order to determine whether this proposed
segregation actually occurs, computer simulations based on a discrete-particle
model for granular material have been carried out in both two and three
dimensions. Most of the results are for the former, which is essentially
equivalent to a very narrow upright container, since considerably smaller
numbers of particles are involved in the calculations.

The granular model employed here is the same as in the horizontal flow
simulations \cite{lev01}; early work with simplified, inelastic, soft-particle
models of this kind appeared in \cite{cun79,wal83}, and a number of similar
models are now in widespread use for modeling the dynamics of granular flow.
The interaction used to provide the spherical shape of the granular particles
has a Lennard-Jones (LJ) form, truncated at a range $r_c$ where the repulsive
force falls to zero. For particles located at $\bm{r}_i$ and $\bm{r}_j$ this is
${\bm{f}}^r_{ij} = (48 \epsilon / r_{ij}) [ (\sigma_{ij} / r_{ij})^{12} - 0.5
(\sigma_{ij} / r_{ij})^6 ] \hat{\bm{r}}_{ij}$ for $r_{ij} < r_c = 2^{1/6}
\sigma_{ij}$; here ${\bm{r}}_{ij} = {\bm{r}}_i - {\bm{r}}_j$ and $\sigma_{ij} =
(\sigma_i + \sigma_j) / 2$. $\sigma_i$ is the approximate diameter of particle
$i$, although because the particles are slightly soft the diameter is not
precisely defined. Normal and transverse damping forces (the latter is
optional) act during the collision (i.e., while $r_{ij} < r_c$); the normal
damping force is ${\bm{f}}^n_{ij} = - \gamma_n (\dot{\bm{r}}_{ij} \cdot
\hat{\bm{r}}_{ij}) \hat{\bm{r}}_{ij}$, and the transverse force is
${\bm{f}}^s_{ij} = - \min (\mu | \bm{f}^r_{ij} + \bm{f}^n_{ij} |, \gamma_s |
\bm{v}^s_{ij} |) \hat{\bm{v}}^s_{ij}$, where ${\bm{v}}^s_{ij}$ is the relative
transverse velocity at the point of contact (which depends on the angular
velocities of the particles and their relative translational velocity). The
value of the static friction coefficient is $\mu = 0.5$, and the normal and
transverse damping coefficients are $\gamma_n = \gamma_s = 5$. Particles are
also subject to a gravitational acceleration $g$.

For convenience the results are expressed in terms of reduced units, in which
length measurements are based on the diameter of the small particles. Such 
units are readily converted to actual physical units. If, for example, the
diameter is actually $10^{-3} m$, then the reduced time unit consistent with
the value $g = 5$ used subsequently corresponds to $0.022 s$, so that the
vibration frequency value $f = 0.4$ employed in these simulations is equivalent
to $\approx 18 Hz$, typical of the values used experimentally. Horizontal flow
rates \cite{lev01} within the strata are then of order $10^{-2} m/s$. 

The particle size distribution is bimodal, with values randomly distributed
over narrow ranges extending from the nominal small and large sizes to values
10\% lower. All particles have the same material density. The sawtooth base
for the two-dimensional simulations is constructed (for reasons of
computational convenience) from a set of grainlike particles positioned in a
manner that describes the desired asymmetric sawtooth profile; the base
particles are forced to oscillate vertically in unison to produce the effect of
a sinusoidally vibrated base, and it is these vibrations that drive the system.
The base particles interact with the freely moving granular particles using the
same force laws as above; in order to ensure reasonably straight sawtooth edges
the diameter of these particles is 0.33 and the distance between their centers
is 0.17. Further discussion of the model appears in \cite{lev01}, and other
more general aspects of the computational methodology are to be found in
\cite{rap95}. Unlike the earlier work, which dealt with a horizontally periodic
system, particles here the are confined to a container with reflecting side
walls; the container is sufficiently high that particles never reach the upper
boundary.

\begin{figure}
\includegraphics[scale=0.55]{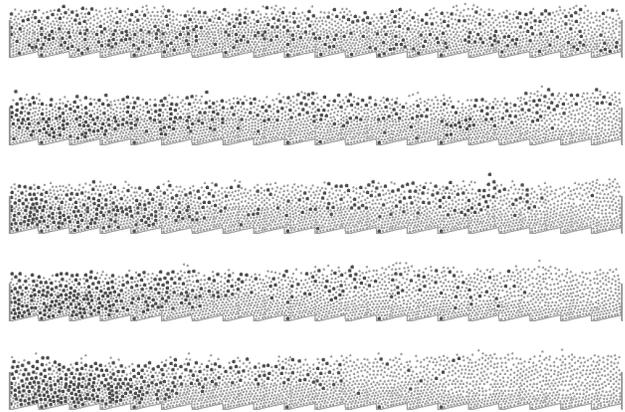}
\caption{Screen images showing stages in the size segregation process for a
two-dimensional system; the large particles are darkly shaded.}
\label{fig1}
\end{figure}

Figure~\ref{fig1} shows a series of screen snapshots (the image resolution is
limited) for a mixture in which there is a 15\% concentration of large
particles with diameter 1.4. The first image corresponds to an early state of
the system, with large particles randomly located throughout the layer.
Subsequent images show the particle distributions after 1000, 2000, 4000 and
16000 vibration cycles; the occurrence of size segregation, in this instance
marked by the leftward motion of the large particles, is clearly visible. 


The presence of the side walls interferes with the stratified horizontal flows
observed with periodic boundaries (corresponding to an effectively infinite
system) \cite{lev01}; nevertheless, in a system of the width considered here,
the stratified flow not too close to the walls is sufficiently strong to drive
the segregation. Any practical implementation of such a separation mechanism
would of course require some means of continuously injecting a granular mixture
at a point above the layer somewhere near its center, and extracting the (at
least partially) segregated products from opposite ends of the container
through suitably located openings.


\begin{figure}
\includegraphics[scale=0.65]{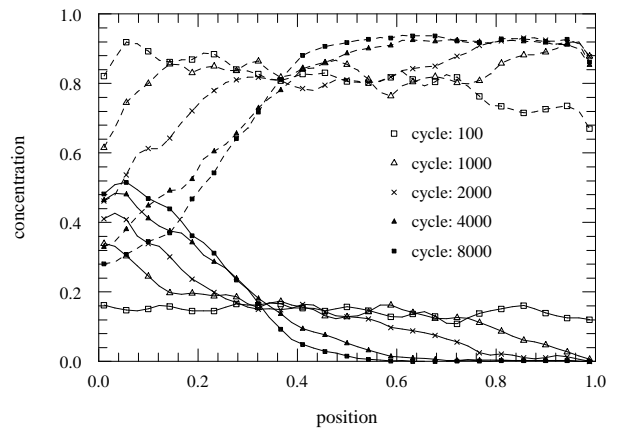}
\caption{Variation of the local concentration of large and small particles
(solid and dashed curves respectively) across the system at several stages
during the segregation process.}
\label{fig2}
\end{figure}

Concentration profiles for the large and small particles at various times are
shown in Figure~\ref{fig2}. The gradual emergence of segregation is apparent,
beginning with an initially uniform system, and finishing with practically all
the large particles having migrated to the left side of the container. The
system width is 180 (in reduced units) for the particular case shown here, and
the nominal number of layers is 8. The frequency is $f = 0.4$ and the amplitude
$A = 1$. The dimensionless acceleration $\Gamma = (2 \pi f)^2 A / g$ is an
important quantity in vibrating systems, with $\Gamma \approx 1$ the minimum
required to excite the layer; a value $\Gamma = 1.26$ is used here, obtained by
setting $g = 5$. The base contains 20 sawteeth of height 2, with an asymmetry
such that the right edge of each tooth is practically vertical. For the case of
granular particles with a unimodal size distribution \cite{lev01} this choice
of parameters produced oppositely directed stratified flows of similar
magnitude (there the system width was just 90, so the relevant velocity profile
is for 10 sawteeth). The large particle size is 1.2 and the concentration 15\%.
Results are averaged over short time intervals and over six independent runs
(with different initial states), and are smoothed to reduce fluctuations.
Similar behavior is observed if the LJ overlap potential used here is replaced
by linear or Hertzian potentials; thus the results do not depend on the choice
of interaction. The fact that segregation occurs for a large particle size of
only 1.2 suggests a highly sensitive mechanism.

In the case of the unimodal size distribution, the stratified flows were
observed \cite{lev01} to depend, often in a complex manner, on the various
parameters used to specify the system. A similar dependence also occurs for
mixtures. Space does not permit a detailed analysis of the dynamical phase
diagram, which would show how the separation rate and direction (as well as the
purity of the segregation products) depend on each of these parameters.
Instead, a few selected examples illustrating the variation with certain key
parameters will be presented, with emphasis on cases where segregation is
strong.

\begin{figure}
\includegraphics[scale=0.65]{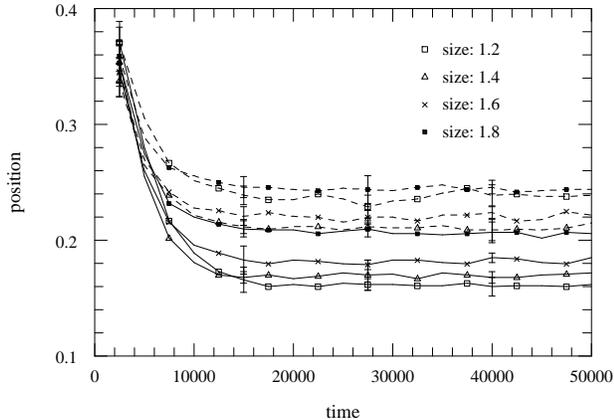}
\caption{Position of the center of mass of the large particles as a function of
time for different particle sizes, and for 20 and 40 sawteeth (solid and dashed
curves).}
\label{fig3}
\end{figure}

Particularly important from a practical perspective is the efficiency of the
segregation mechanism for different large particle sizes. Figure~\ref{fig3}
shows this size-dependence for two sawtooth widths. The degree of segregation
is expressed in terms of the time-dependent position of the center of mass of
the large particles; the occasional particle trapped in a sawtooth groove
before its migration is complete will adversely affect such a measurement, as
will any other effects preventing the formation of a well-defined vertical
boundary between the two segregated species (see below). The results are again
averaged over six independent runs. The final value depends on the large
particle size; in the case of 20 sawteeth it increases monotonically with size,
reflecting the increasing space occupied by the large particles, but for 40
sawteeth the results for diameter 1.2 deviate from the expected sequence. The
times required to reach the limiting position are essentially the same.


\begin{figure}
\includegraphics[scale=0.55]{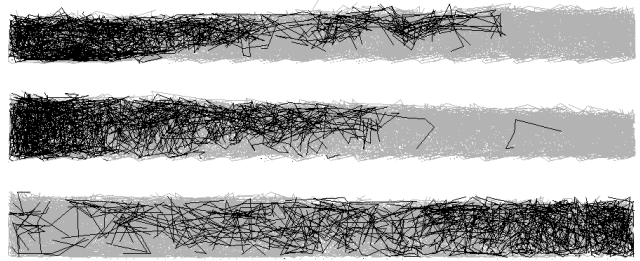}
\caption{Trajectories of 2\% of the particles after segregation has ended
(large particle trajectories are in black), for pairs of values of large
particle size and number of sawteeth (1.4, 20), (1.2, 20), and (1.2, 80).}
\label{fig4}
\end{figure}

While a detailed picture of the particle movement over the entire history of
the segregation is not readily obtained because of the strong degree of mixing
that occurs, a study of the trajectories for a subset of particles, once the
process has effectively ended, demonstrates where the particles tend to be
located. Figure~\ref{fig4} shows short trajectory segments for just 2\% of the
particles after 35000 vibration cycles; each trajectory consists of five points
at 100 cycle intervals. In the first two examples (with 20 relatively wide
sawteeth, and large particle sizes 1.4 and 1.2) the large particles congregate
on the left; they occupy a region that includes the section of the layer
abutting the wall extending from top to bottom, and adjacent to this a roughly
wedge-shaped region that tapers upwards to the right. The non-rectangular shape
of this region explains why the center of mass position does not fully
characterize the final state. In the third example (80 very narrow sawteeth,
particle size 1.2) segregation occurs in the opposite direction and is less
complete.

In those cases where large particles migrate to the left, the shape is a
consequence of a dynamic equilibrium involving both stratified flow, in which
particles in the upper levels are driven horizontally to the left by the
sawteeth and lower particles to the right, and Brazil-nut ``buoyancy'' which
raises the large particles. In a practical implementation of such a segregation
apparatus, large particles would be removed near the upper left corner and
small particle via an opening near the lower right. The case where large
particles migrate to the right reflects the absence of leftward stratified flow
due to the sawtooth base. In all cases shown here, irrespective of direction,
the horizontally driven large particles eventually displace the small particles
from the appropriate side wall.


The corresponding three-dimensional version of this system consists of a
rectangular base with linear sawtooth-like grooves; similar results would be
expected from simulations of such a system. Since it is more interesting to
consider a problem involving an alternative geometry, a system with circular
symmetry is described here. 

In this three-dimensional example the base consists of a set of concentric
circular grooves, shaped to produce a radial profile identical to that used in
two dimensions. Interactions between particles and the base are handled with a
simple extension of the approach used above, by employing a series of
concentric donut-like base ``particles'' to represent the grooves (only the
radial coordinates of such particles enter into the force computation). Since a
substantially larger number of granular particles are necessary to achieve a
linear size similar to that of the two-dimensional system, only a limited
exploration of this problem has been undertaken. The transverse damping force
is omitted here; while its absence does not seriously influence the observed
behavior, indicating once again that the phenomenon is not especially sensitive
to the details of the model, it does reduce the computational effort. 

\begin{figure}
\includegraphics[scale=0.4]{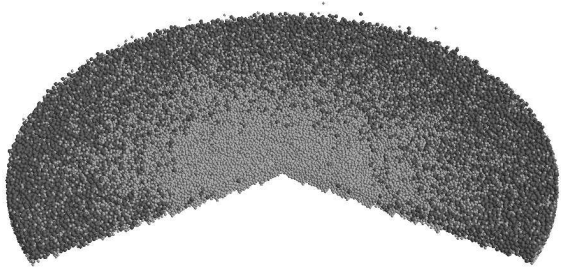}
\caption{Segregation in a three-dimensional system; a wedge-shaped region has
been removed from the front and the container is not shown (large particles are
darker).}
\label{fig5}
\end{figure}

Figure~\ref{fig5} is a screen image from one such simulation showing the system
after 2000 vibration cycles. The container diameter used here is 180 and the
nominal layer thickness is 8; there are 11 concentric sawtooth grooves, and the
large particles have size 1.5 and concentration 33\%. The sawtooth profile,
whose influence is apparent in the positioning of the particles closest to the
base, is oriented so that large particles should migrate towards the outer
boundary of the cylindrical container while small particles move inwards, which
indeed is exactly what is seen to occur.


In summary, computer simulations employing a type of granular model whose
viability has been established in other studies of grain flow have been used to
investigate a novel mechanism for achieving size segregation. Selected results
have been presented here, with a more detailed determination of the parameter
dependence to be published in due course. Since separation and sorting are
essential functions in the processing of bulk granular materials, such an
approach, implemented for example as a continuous flow device, has potential
industrial value. It remains to be seen whether the behavior of real granular
matter is in accord with the predictions of these simulations.

\begin{acknowledgments}

This research was supported in part by the Israel Science Foundation. 

\end{acknowledgments}

\bibliography{horizseg}

\end{document}